\documentstyle[floats,psfig,aps,twocolumn]{revtex}

\def\kA{\mbox{${\bf k}_A$}} 
\def\kB{\mbox{${\bf k}_B$}}
\def\ka{\mbox{${\bf k}_1$}} 
\def\kb{\mbox{${\bf k}_2$}}
\def\pa{\mbox{${\bf p}_1$}} 
\def\pb{\mbox{${\bf p}_2$}}
\def\pc{\mbox{${\bf p}_3$}} 
\def\pd{\mbox{${\bf p}_4$}}
\def\pe{\mbox{${\bf p}_5$}} 
\def\pf{\mbox{${\bf p}_6$}}
\def\rA{\mbox{${\bf r}_A$}} 
\def\rB{\mbox{${\bf r}_B$}}

\def\qa{\mbox{${\bf q}_1$}} 
\def\qb{\mbox{${\bf q}_2$}}
 
\def\ks{\mbox{${\bf k}_S$}}
\def\r{\mbox{${\bf r}$}} 
\def\rs{\mbox{${\bf r}_S$}} 
\def\K{\mbox{${\bf K}$}} 
\def\d{\mbox{${t/m}$}} 
\def\k{\mbox{${\bf k}$}}
\def\q{\mbox{${\bf q}$}}
 
\begin{document}
\draft
\title{HBT correlation functions from event generators -  a reliable approach
 to determine the size of the emitting source in ultrarelativistic heavy ion 
 collisions ?}

\author{F. Gastineau and J. Aichelin}

\address{SUBATECH \\
Laboratoire de Physique Subatomique et des Technologies Associ\'ees\\
UMR Universit\'e de Nantes, IN2P3/CNRS, Ecole des Mines de Nantes\\
4, rue Alfred Kastler,
F-44070 Nantes Cedex 03, France.}
\date{\today}
\author{\begin{quote}
\begin{abstract}
Employing neXus, one of the most recent simulation programs for heavy ion 
collisions, we investigate in detail the Hanbury-Brown Twiss correlation 
function of charged pions and kaons for central reactions 158 AGeV Pb+ Pb. 
For this study we supplement the standard simulation program by electromagnetic
interactions. We find a wide distribution of freeze out times which corresponds
to a broad distribution of source sizes.
Strong space-momentum space correlations as well as the electromagnetic
interaction between {\bf the correlated pairs and the environment}, which is of
the same size as that between the hadrons of the correlated pair, modify
strongly the correlation function. 
It can still be well parameterized by $C({\bf k })= 1+\lambda e^{-{4\bf
k}^2R^2}$, where k is the relative momentum.  $R$ is,
however, not a simple function of the source radius but the complicated
space-time structure of the source as well as the space-momentum space 
correlations and the final state Coulomb interactions are encoded. 
The true source radius, as given by the rms radius of the emission points,
and R differ by up to $30\%$.   
\end{abstract}
\end{quote}}

\maketitle

\section{Introduction}

Which particle and energy densities are reached during an ultra relativistic heavy
ion reaction? This is one of the present key questions in this field.  If the
density exceeds a critical value, theory predicts that a plasma, consisting of
quarks and gluons, is produced whereas below that density the matter stays in its
hadronic phase, consisting of mesons and baryons.

Theoretical predictions on the density and the energy density obtained in present
(SPS) or future (RHIC and LHC) heavy ion reactions are rather vague because
the understanding of the initial phase in which projectile and target become
decelerated is still a theoretical challenge without a reliable answer yet.
Therefore the present efforts concentrate on an experimental determination of
these densities.

It is all but easy to measure coordinate space dependent quantities in heavy ion
reactions. Single particle spectra contain direct information on the
momentum space only. Therefore more complicated observables have to be employed to
study the density. Probably the most powerful but certainly the most
employed approach is based on a suggestion of Goldhaber, Goldhaber, Lee and
Pais (GGLP) \cite{GGLP} who used the Hanbury-Brown and Twiss (HBT) effect
 \cite{HBT56}. It is based on the fact that the amplitudes of two indistinguishable processes 
interfere. If an object emits two identical bosons or fermions with 
momenta $\ka$ 
and $\kb$
which are registered by two detectors the dependence of
the cross section ${d^6\sigma\over d^3k_1d^3k_2}$ 
on the relative distance between the detectors
or on the relative momentum of the registered particles is  an image of the 
emitting source. This method has been proven to be quite powerful for measuring
the size of stars \cite{HBT56}.

In heavy ion physics the situation is much more difficult:\\
a) There is nothing like a static source which emits particles.
The system has a complicated geometrical structure and expands in space and 
time.  Therefore 
not only one parameter (like in astrophysics the rms radius of a star) but the 
functional dependence of several parameters on time is encoded in the
data.\\ 
b) There is no unique definition of when a particle is emitted because the emission
is in reality a disconnection from the expanding system. Usually one assumes
that after its last collision the particle can be considered as emitted. 
The space-time point of the last collision is called freeze out point.\\ 
c) There are long range Coulomb forces which change the particle momenta
after emission.\\ 

Two approaches are possible in this situation:\\ 
A) One neglects the
complications, limits the problems with the geometry by using only particles
which are emitted at midrapidity and parameterizes the source with a couple of
parameters. The parameters are then determined from the measured one and two 
particle cross sections \cite{ccs}. For a review on the present status of this approach we refer
to the recent paper of Wiedemann and Heinz \cite{HEINZ}.
The problem is that it is not easy to reveal 
the physical significance of the numerical values of these parameters.\\ 
B) One employs numerical simulation programs which have been validated
by reproducing the measured single particle spectra. These programs, however,
are semiclassical in their nature and hence particles and not wave functions
are propagated. Therefore one has to extend these programs by a quantum
calculation (classical particles do not interfere) if one would like to employ
them for the prediction of the HBT correlation
function. Here the rationale is as follows: If the extended program
reproduces the correlation data, one can assume that it describes properly not
only the momentum but also the coordinate space distribution of the particles. Hence
the distribution of the particles in these simulation programs 
can be used to calculate densities and energy densities. 
The first suggestion to use the simulation programs to study the HBT effect has
been advanced in \cite{gy}. Later Cs\"org\"o et al. \cite{cso} and Pratt et al.
\cite{pr1} have performed the first calculations using Fritjof, an event simulation
program. In
these calculations it has been assumed that a) the Coulomb interaction can 
simply be taken into account by multiplying the two particle Wigner density by 
the Gamow correction factor and b) one can identify the Wigner density of the nucleon
with the classical phase space density. The latter is obtained by averaging the phase
space points of the particles at the end of the simulation over many events.
This has become the standard approach for a comparison of simulations and
experiments \cite{sul}. Already in ref. \cite{gb} it
has been found that in heavy ion collisions the Gamow correction factor 
overpredicts the true effects and rather
a classical trajectory calculation has to be employed. Recently several
propositions have been made to improve the treatment of the Coulomb interaction
\cite{al,a,b,ac}. The difference between
the classical one body phase space density and the Wigner density of a emitted
particle has been pointed out in \cite{AIC}.

Here we study these both assumptions. In particular,
using one of the up to date simulation programs which reproduces the 
single particle spectra quite well (section 2), we study how the source 
parameters, obtained from the 
asymptotically observed correlation function, are related to the true 
source. This includes a study of the source (section 3), a study of the 
Wigner density of
the emitted particles and its relation to the classical phase space density
(section 4), a
study of the influence of the Coulomb fields (chapter 5),  a study of the influence 
of the resonance decay and a study of the kaon correlation function. 
We will show that the relation between the source and
its image (coded in the correlation function) is all but simple and that
the source density as determined by the correlation function may differ by a
factor of two from that determined directly in the simulation program.

Whereas the study of the difference between the Wigner density and 
the classical phase space density and the consequences for the 
interpretation of the correlation function in terms of a source
radius is rigorous, the study of the influence of
the Coulomb interaction can be only considered as a first step. It shows the
order of magnitude of the corrections one has to expect but present computer
power does not allow to supplement the simulation programs by the solution of a 
Schr\"odinger equation to calculate the (anti)symmetrized wave function of the
correlated pions in the field of all other charges.  
  
\section{The method}
\subsection{neXus}
To study the space time structure of ultrarelativistic heavy ion collisions one
has to rely on simulation programs like VENUS \cite{WER93}, 
neXus\cite{W1} or URQMD \cite{URQ98}. These event
generators describe the reaction from the initial separation of projectile and
target until the final state which is registered in the detectors. They
assume that during the heavy ion reaction strings are formed which 
disintegrate into hadrons. The hadrons are treated semiclassically and can scatter 
among themselves elastically or inelastically before they are detected. 
For our study we use neXus (version 2.0 beta), the successor of VENUS, one of the most developed event
generators, which describes the single particle spectra of the 
observed mesons and baryons  at CERN energies quite well. We refrain from  
describing the details of this model and refer to ref. \cite{W1}. The
program is used in its version without droplet formation in order to avoid the
unknown Coulomb forces between droplets and hadrons. 
In using one of 
these models we follow here the common believe that up to the last strong
interaction collision 
(freeze out) the hadrons are well described by a semiclassical approach.

\subsection{Implementation of the electromagnetic interaction in neXus}
 In order to study more rigorously the influence of the electromagnetic interaction on the 
 two-particle correlation function we have to implement it directly into the
 simulation program. Because the average distance $r(t_2)$ (where 
 $t_2$ is the time when the later emitted hadron has its last collision)
 between the correlated hadrons  is rather large (as we will see in part
 III)
 it is sufficient to implement the interaction on a semiclassical level
 \cite{gb,pratt}. Correlated hadrons or pairs we call those pairs of 
 particles which 
 have a relative momentum smaller than 100 MeV. As we will see later
 the correlation function differs from one only for
 those pairs and hence only they 
 carry information on the size of the system. 
 
 The electromagnetic interaction of a particle $i$ with  another charged
 particle $j$ is given by 
\begin{equation}\label{inter}
 m_i\frac{d u_i}{d \tau} =\sum_{j\neq i} \frac{e_i}{c}F_{ij}^{\mu \nu} u_i^{\nu}.
 \end{equation}
Here j runs over all charged particles of the system. 
The particle $i$ has the mass     $m_i$,  the charge $e_i$ and the 
4-velocity $u_i=\{1, {\bf v} \}/\sqrt{1- {\bf v} ^2 } $. 
The tensor $F_{ij}^{\mu \nu }$  is given by
\begin{equation}\label{fmn}
F_{ij}^{\mu \nu } = \frac{e_j}{c} \frac{X^{\mu}u_j^{\nu} - X^{\nu}u_j^{\mu}}{
( \frac{1}{c^2} (u^{\lambda}_j X^{\lambda})^2 - X_{\lambda} X^{\lambda } )}
\end{equation}
where $X^{\lambda}$ is the relative 4 - distance between the particles $i$ and $j$. \\  

The formula ~(\ref{inter}) assumes point like particles and the absence of
acceleration of the particle j. In our case this is a valid assumption because at high energies
the electromagnetic force changes the momenta only
little. Applying this formalism at each time step in the neXus program 
(which updates the particle positions and momenta at common light cone times) 
we create Coulomb trajectories of all charged particles.

\begin{figure}[htbp]
\centerline{\psfig{figure=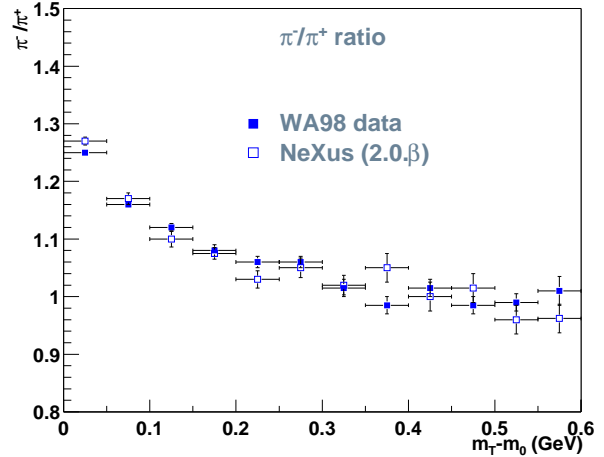,width=8.6cm}}
\caption{$\pi^-$ spectrum divided by the $\pi^+$ spectrum, both obtained by the 
WA98 collaboration [11], as compared to the NeXus result.  }
\label{pipi}
\end{figure}

For the analysis presented later we store the mutual Coulomb force between all
charged particles at all time steps. Whether two mesons form a 
correlated pair we know only at the end of the simulation. 
Using these stored forces we can separate the change of the relative momentum
of the pair particles $\k_{S}$ due to their mutual interaction 
($ \Delta  \k_{pair}$) from that due to their interaction with the 
environment (i.e. all other charged particles)($\Delta \k_{environ}$). 
The final relative momentum of the correlated pair $ \k_{final}$ is 
\begin{equation}
\k_{S} + \Delta\k_{environ} +\Delta\k_{pair} = 
\k_{final}. 
\end{equation}
As we will see in this classical trajectory calculation the average change of
the relative momentum of the correlated pair particles due to their mutual 
Coulomb interaction is about as large as that due to the interaction with the environment. This
suggests that also in a quantal calculation both are equally important. The
importance of the Coulomb interaction between a particle and its environment 
can also be seen experimentally by comparing the measured $\pi^+$ and 
$\pi^-$ spectra as done in fig.\ref{pipi}. 
We see that at relative momenta smaller than 100 MeV the spectra differ 
considerably and that this difference is well reproduced by the neXus 
calculation including the classical Coulomb interaction. This relative momentum
has to be compared with the relative momentum $\k$ for which the correlation 
function $C(\k)$ becomes one and hence ceases to carry information on the 
source size. As we will see later $\k$ is as well around 100 MeV. Hence the
influence of the Coulomb interaction on the correlation function cannot be 
neglected.   

\section{Space time structure of the source}
\begin{figure}[htbp]
\centerline{\psfig{figure=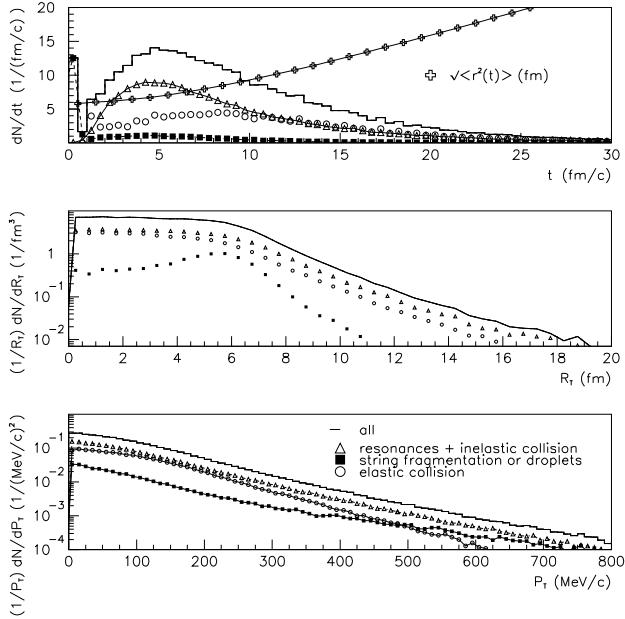,width=8.6cm}}

\caption{Distribution of freeze out times, freeze out radii and freeze out 
momenta for charged $\pi$'s with $y_{cm}-1<y<y_{cm}+1$ from neXus+Coulomb. 
Black squares show particles from string fragmentation, open
circles show pions from elastic collisions and open triangles those
from resonance decay or from inelastic collisions.}
\label{avdrogene}
\end{figure}

The first question we address is when and where the particles, which are finally seen 
in the detectors, are created. We limit ourselves to pions
emitted at
midrapidity $ y_{cm}-1 \le y \le y_{cm}+1$. For our study we use the reaction
Pb + Pb at 158 AGeV for b $ \le $ 3.2 fm. 
Fig.\ref{avdrogene} top presents the distribution of freeze out 
times of charged mesons.
We display this time separately for particles created 
by string decay,
by resonance decay or inelastic collisions, and by elastic collisions. We see
no sudden freeze out, rather a distribution which is 
almost flat between 3 and 10 fm/c. Without rescattering we would
have seen one peak at 1 fm/c (string breaking) and another one around 5 fm/c
($\rho$ decay ).

The distribution of transverse distances $R_T = \sqrt{x^2+y^2}$ of pions
at freeze out with respect to the center of the reaction is displayed in the
middle part of the figure. We observe as well a rather flat distribution followed by a
very long tail which ranges up to 18 fm! Hence according to the simulation program
rescattering made it impossible to define a unique radius of the system at freeze
out which could hopefully be measured. 
The transverse momentum distribution 
of the pions does not strongly depend on their origin. This can be concluded
from the bottom part of fig.\ref{avdrogene}, where the momentum distribution of
the pions is displayed.

More interesting than the average properties of all particles are that 
of those pions, which finally have a small relative momentum
with respect to another pion of the same charge state. We will see later that 
the correlation
function differs from 1 only for those pairs of particles which have a relative
momentum smaller than 100 MeV. These pairs we call  correlated pairs and only
they carry information on the size of the system. 

\begin{figure}[htbp]
\centerline{\psfig{figure=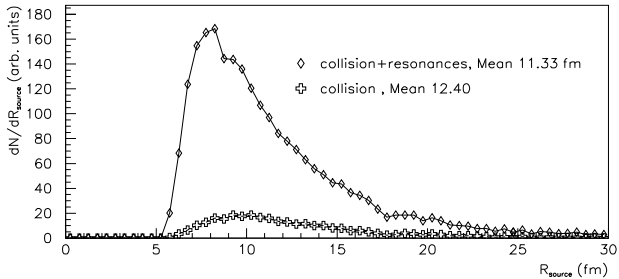,width=8.6cm}}
\caption{Distribution of source sizes for pairs with a relative momentum of less
than 100 MeV.}
\label{rsyst}
\end{figure}

The distribution of freeze out times $t_2$ of those pairs can be directly translated into a
distribution of source sizes $R_{source}(t_2) = 
\sqrt{{1 \over N_{pion}}\sum_{all\ pions}{\bf r}_i(t_2)^2}$. 
This distribution of source sizes is displayed in fig.\ref{rsyst} for two cases:
for all pairs and for those pairs where none of the pions come from
resonance decay. We see a broad distribution with a maximum at 8 (10)fm and a 
mean value of 11.3 (12.4)fm. For
calculating the mean value we have omitted all particles beyond $R_{source}$ =
30 fm, i.e. pions which come from long living resonances like $\omega$ or 
$\Lambda$.

Thus there is neither a common freeze out time nor a common freeze out volume.

\begin{figure}[htbp]
\centerline{\psfig{figure=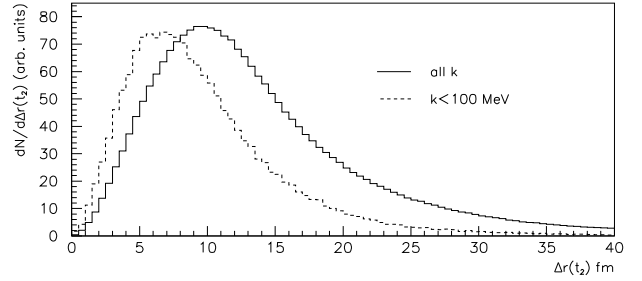,width=8.6cm}}

\caption{Distribution of relative distances at freeze out for pairs with a final momentum 
of $k < 100 MeV$ in comparison with that of all pairs of positively charged pions.}
\label{relat}
\end{figure}

In fig.\ref{relat} we compare the distribution of the relative
distance $r(t_2)$ at freeze out for the correlated pairs in comparison with
that for all pion pairs.  First of all, the average distance is above 10 fm.
We observe furthermore that the average value 
for the small relative momentum pairs  ($\sqrt{<r^2(t_2)>}$ = 10.8 fm) 
differs considerably from that for all pairs ($\sqrt{<r^2(t_2)>}$ = 15.6 fm).
Consequently, the pairs with a small relative momentum are not representative
for all pairs. This one has to keep in mind if one likes
to interpret the correlation function in terms of physical source radii.

\begin{figure}[htbp]
\centerline{\psfig{figure=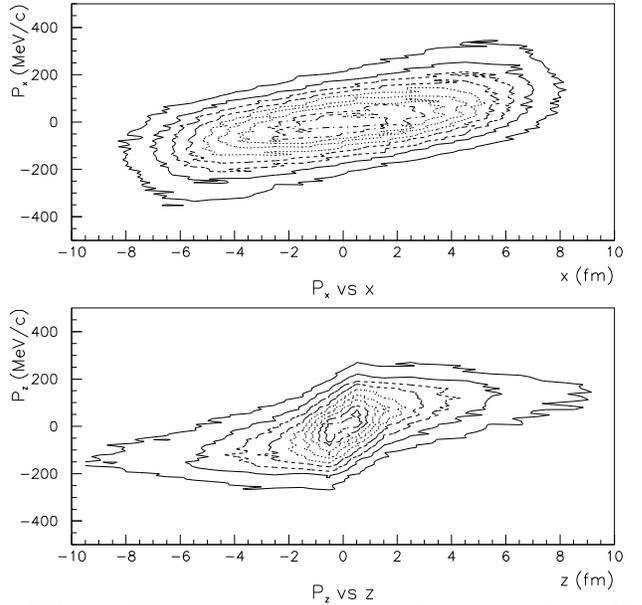,width=8.6cm}}
\caption{Momentum-position correlation in the neXus model.}
\label{imp-pos}
\end{figure}

The difference between the distribution of relative distances of
pairs with $k<100MeV$ and that of all pairs is a consequence of strong 
space - momentum space correlations, in beam (z) direction as well as in the
direction
of the impact parameter (x). These correlations of freeze out momenta and
positions are displayed in fig. \ref{imp-pos} and are well known. 
They give rise to the identity of rapidity and space time
rapidity. We see that even in the small rapidity interval around
midrapidity, in which we perform our investigation, these correlations 
are by far not negligible.

Now, after we have described the source, we can investigate to which extend its
properties can be recovered by measuring correlation functions.

\section{The correlation function without electromagnetic interactions}
\subsection{The formalism}
 All semiclassical simulation programs propagate classical particles which 
have a sharp
momentum \kA and a sharp position \rA. This is conceptually necessary in order
to determine the sequence of collisions and the center of mass energy of each
collision. On the other hand, two particles emitted with sharp momenta and 
sharp positions do not interfere because their trajectories are distinguishable.
In order to model physics which is based on the interference of wave functions, 
like the HBT effect, one has to treat the particles as quantum particles for which the
uncertainty principle is fulfilled. Therefore a transition from classical
to quantal physics is necessary. In view of the construction of the simulation
programs this transition can only be made after the last collision.
 
This transition from classical to quantal physics requires more than 
the knowledge of \kA and \rA \ provided by the simulation programs. 
In order to construct a wave function one needs at least one additional
parameter (the width of the wave function) which is not provided by the
simulation program. A popular  
choice for the wave function is a Gaussian superposition of plane waves. 
This wave function requires 
only one unknown parameter, the variance L/4, and there is even hope that its
physical relevance can be understood, as we will see.

Two identical particles emitted at $t_1$ and $t_3$ at the source points \rA \ and \rB \ with the
momenta centered around \kA \ and \kB  \ observed in detectors with the momenta 
\ka \ and \kb \ can have two indistinguable
trajectories. Either the particle emitted at \rA \ arrives with momentum \ka \ 
(and that from \rB \ with momentum \kb)
\begin{eqnarray}  
A_1 &=& \int d\pa d\pb <\ka,t_2 |t_1, \pa><\pa,t_1 |\Phi_A> \nonumber\\
    &&\cdot <\kb, t_4 |t_3,\pb><\pb,t_3 |\Phi_B>
\label{a1}      
\end{eqnarray}
or that emitted at \rA \ arrives with momentum \kb \ and
correspondingly that emitted at \rB \ with \ka
\begin{eqnarray}  
A_2 &=& \int d\pa d\pb <\ka,t_2 |t_1, \pa><\pa,t_1 |\Phi_B> \nonumber \\
    &&\cdot <\kb, t_4 |t_3,\pb><\pb,t_3 |\Phi_A>  . 
\label{a2}
\end{eqnarray}
For the wave function of the emitted particle
\begin{equation}
<\pa,t_1|\Phi_A>= f_A(\pa)e^{i\pa \rA}e^{-i\pa^2t_1/2m} 
\end{equation}
we assume a Gaussian form
\begin{equation}
f_A(\pa)=({L\over 2\pi})^{3/4} e^{-(\pa-\kA)^2L/4} .
\end{equation}
If L is large the momentum of the particle is close to the source momentum but
we loose the information on the source position. If L is small we have the
opposite scenario.

$<\ka,t_2 |t_1, \pa>= \delta^{(3)}(\ka-\pa)e^{-i\pa^2(t_2-t_1)/2m}$ is the propagator 
in momentum space. 
Because both trajectories are indistinguishable their amplitudes
have to be added to obtain the correlation function
\begin{eqnarray}
\lefteqn{ C^{(2)}(\k)={\int d\K |A_1+A_2|^2 \over  \int d\K (A_1^2+A_2^2)}}\\
&=& {  g(\k-\ks)+g(\k+\ks)+2g(\k)g(\ks) cos(2\k\rs) \over g(\k-\ks)+g(\k+\ks)}
\end{eqnarray} 
where $\k = {\ka - \kb \over 2}\ , \K = (\ka + \kb) \ ,\ks = {\kA - \kB \over 2} 
\ ,\rs = (\rA- \rB)$ and $g(\k)= ({L\over \pi})^{3/4} e^{-\k^2L}$.
In our calculation we update the coordinates of the first emitted particle
until the emission of the second particle of the correlated pair (at $t_2$).
Therefore $cos(k_\mu r^\mu) = cos \k \r$ where $r_\mu = \{0,
\r(t_2)\}$   and the time component of the 4-vector of the relative
distance disappears from the correlation function.
If the two bosons have the same source momentum (\kA=\kB) and are
measured in the detector with the same momentum we obtain
an enhancement of $C^{(2)}(\k)$ by a factor of two as compared to
distinguishable particles. However, if the source momenta are different, the
result is more complicated. 

If L is sufficiently small, the difference
in the arguments of the exponential functions can be neglected and we obtain 
\begin{equation}
C^{(2)}_{SM}(\k)= {g(\k-\ks)(1+e^{-\k^2L} cos(2\k\rs))\over g(\k-\ks)}.
\end{equation}
This approximation is called smoothness assumption. It is frequently used in
the literature for the following reason: The measured correlation function can be well
described by a Gaussian function \cite{AIC} 
\begin{equation}
C({\k}) = 1 + \lambda exp(-{4R^2}\cdot \k^{2}) 
\label{nneu}
\end{equation}
where $\lambda$ allows for the correction of a possible coherence in the source 
or for pairs of non identical particles. Assuming that the emission points 
have as well a 
Gaussian distribution in coordinate and momentum space 
\begin{equation}
S(\ks,\rs)\propto
e^{-\ks^2B}
e^{-\rs^2/C}, 
\label{a15}
\end{equation}
where \ks \ and \rs \ are the relative distances in momentum and coordinate space,
the convolution of  
$C^{(2)}_{SM}(\k)$ with the source distribution function S(\ks,\rs) yields
\begin{eqnarray}
&\;& C_{SM}(\k) = \nonumber \\    
&\;&{\int d\rs d\ks S(\ks,\rs)g(\k-\ks)(1+ e^{-\k^2L} cos(2\k\rs)
 \over 
\int d\rs d\ks S(\ks,\rs)g(\k-\ks)} \nonumber \\
&\;&=1+e^{-\k^2(L+C)}.\\ 
\label{a6} 
\end{eqnarray}
 The apparent source size ${4R^2}$ is, 
 in the Gaussian source approximation,
four times the sum of the variance of the source C/4 and of that of 
the wave function squared
L/4. In this approximation we find the following relation between
the root mean square radius of the freeze out points $\sqrt{<r(t_2)^2>}$ 
and the apparent source radius
\begin{equation}
 <r(t_2)^2> = 3R^2 = 3/4(L+C).
\label{neun}
\end{equation}
Thus in the smoothness approximation one has only to fit
the experimentally observed correlation function
by a function of the form of eq.\ref{nneu} and can then use eq. \ref{neun} 
to determine the desired $<r(t_2)^2>$. Consequently one can avoid to 
perform a Fourier transform, 
a difficult task in view of the large experimental error bars. 
In addition this equation is easy to interpret.

In the simulation programs we do not have a continuous source distribution.
Rather we have to sum over all emission points:
\begin{equation}  
C(\k)={\sum \int d\K |A_1^i+A_2^i|^2 \over \sum \int d\K (|A_1^i|^2+|A_2^i|^2)}
\label{a8}
\end{equation} 
which yields in the smoothness approximation \\
\begin{eqnarray}
& \;& C_{SM}(\k)  =  \nonumber \\  
 &\; &{\sum_{s=1}^{N_{ev}}\sum_{i=1}^{N_{pairs}} 
(1+e^{-\k^2L}cos(2\k\rs_{s,i}))*g(\k-\ks_{s,i}) \over 
\sum_{s=1}^{N_{ev}}\sum_{i=1}^{N_{pairs}} g(\k-\ks_{s,i})}
.
\label{a7} 
\end{eqnarray}
$N_{pairs}$ is the number of pairs in the event s.

Now it is important to realize which information is needed to calculate
a correlation function and which information is to our disposal in these simulation
programs. As seen in eqs. \ref{a1},\ref{a2},\ref{a8},\ref{a7} we need\\ 
a) the single particle wave function for the two correlated particles
emitted from emission points  with relative coordinates $\rs$ and $\ks$.\\
b) the distribution of emission points. Because in the derivation
of the correlation function \cite{AIC} the source is treated classically 
these emission points $\rs, \ks$ 
are either given by a source function like eq.\ref{a15} or as a sum of delta 
functions in
coordinate and momentum space as in eq.\ref{a8},\ref{a7}.\\    
The simulation programs treat the rescattering of the hadrons
like the scattering of classical billiard balls. Only initially, when the hadrons
are created by string decay, quantum mechanical constraints are taken into
account. Consequently, the simulation programs present the time evolution 
of a classical chaotic system whose initial configuration is inspired 
by quantum mechanics. Hence we have to our disposal the classical n-body phase
space density. Assuming that emission takes place at freeze out the classical 
n-body phase space density allows to calculate the emission points.
 \begin{figure}[htbp]
\centerline{\psfig{figure=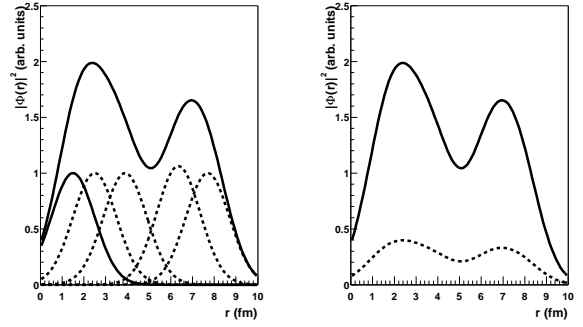,width=8.6cm}}
\caption{2 different sets of single particle wave functions squared 
$\sum_i \mid \Phi_i(r)
\mid ^2$ which give the same one body Wigner density $\int F_W^1(r,k)dk$.
On the right hand side $\mid \Phi_i(r)\mid ^2 = (\int F_W^1(r,k)dk)/5$.} 
\label{wig}
\end{figure}

It has been argued in the past \cite{sc,zhang} that the knowledge of  the 
classical n-body phase space density is sufficient to determine
the single particle wave function as well.  The argument goes as follows: 
From the classical n-body phase space density we can construct the 
classical one body phase space density
(we use the nonrelativistic formulas here because there the
argument becomes more transparent)
 \begin{equation}
F^1_{cl}(\r,\k,t)= \sum_{i=1}^n \delta^{(3)}(\k -
\k_i(t)) \ \delta^{(3)}(\r
- \r_i(t)) .
\label{a8b}
\end{equation}
Averaged over many events\cite{sc} or over one event \cite{zhang}
the classical one body density can be identified with
the quantal one body Wigner density 
 \begin{equation}
F^1_{W}(\r,\k,t)= {1\over N_{ev}}\sum_{N_{ev}}\sum_{i=1}^n g(\k ,\r ,\k_i(t), \r_i(t))
\label{a4}
\end{equation}
where $g(\k ,\r ,\k_i(t), \r_i(t))$ is the Wigner density of the single particle
wave function $\Phi(\k ,\k_i(t),\r_i(t))$ defined as
\begin{eqnarray}
g(\k,\r,\k_i(t),\r_i(t))&=&\int {d\q\over (2\pi)^3} e^{i\q\r} 
\Phi^*(\k+\q/2 ,\k_i(t),\r_i(t))\nonumber \\ 
&\cdot&\Phi(\k-\q/2 ,\k_i(t),\r_i(t)) .
\label{a3}
\end{eqnarray}
Now one assumes that all $\k_i(t),\r_i(t)$ are identical. Then we can identify
\begin{equation}
F^1_{W}(\r,\k,t) = n\cdot g(\k ,\r)
\label{a9} 
\end{equation}
Knowing $g(\k ,\r)$ one can 
calculate the amplitudes $A_1$ (eq.\ref{a1}) and $A_2$ (eq.\ref{a2}). 

There are, however, three shortcomings in this argument:\\
1) There is an infinity of single particle wave functions $\Phi$(eq.\ref{a3}) 
which give the same one body  Wigner density (eq.\ref{a4}). They yield, however,
quite different correlation functions. Fig.\ref{wig} shows 
as example two sets of single particle wave functions which give the same 
one body Wigner density but different two body correlations C({\bf k}). 
Thus eq. \ref{a9} is {\bf one out of an infinity of choices} of a single
particle wave function for a given one body  Wigner density. 
One may wonder if this is a good choice because it ignores all 
correlations between the $\k_i(t)'s$ and $\r_i(t)'s$. In addition
it is difficult to motivate why until freeze out the particles are 
treated as point like in coordinate and momentum space whereas 
thereafter they are well described by a wave 
function for which $\Delta k \Delta r >> \hbar$. It is more natural to
replace the precise momenta and positions of the particles by the "most
classical" wave functions around the classical $\k_i(t)'s$ and $\r_i(t)'s$, 
i.e. Gaussians which fulfill the uncertainty principle.
In general, the ignorance of L and the impossibility to derive from the known one body 
Wigner density the single particle wave functions are two ways to express
the same unknown physics. \\

Averaging over many events raises even more questions:\\ 
2) Two particles emitted next to each other (i.e. those which are finally interfering
strongly) have a mutual Coulomb repulsion, which introduces a two body
correlation between the pair even on the classical level. Averaging over many
events (which means to allow that particles from different events interfere)
destroys this correlation. Thus event averaging destroys correlations
already present on the classical level. \\
3) There is no proof that the classical n-body phase space density averaged 
over many events is equivalent to the quantal one body Wigner density. On the
contrary there are good reasons that this is not the case, if the size of the
system is comparable to the width of the single particle wave function:  
If all particles are sitting on top of each other a classical n-body density
is point like whereas the Wigner density has to obey the uncertainty principle.

For the determination of the correlation function from experimental data the
denominator is obtained by event mixing. For our theoretical studies this is 
not necessary. Therefore we would not like to comment here on the consequences
of the event mixing procedure for the correlation function. Numerically
we observe quite a difference in the results if we replace the denominator of
eqs.\ref{a8},\ref{a7} by the corresponding quantity for mixed events. This is
caused by mutual Coulomb interactions (see 2)) which get lost by event mixing.

\subsection{The Correlation Function in neXus without electromagnetic interactions}
As discussed, for a Gaussian source (eq.\ref{a15}) with no 
space-momentum space correlations there is a simple relation between the
true source radius and the apparent source radius (eq.\ref{neun}).

{ Is this observation also true for the source as given by the simulation 
programs? This is the seminal question which we will discuss in this and in 
the next paragraphs. In the simulation programs
we are confronted with three different source radii
\begin{itemize}{
\item the {\bf classical source radius $R_{class} = \sqrt{<r(t_2)^2}>$} as 
given by the simulation  program where $<..>$ means averaging over all pions
which are part of a pair. Their radius is taken at the time when the
later emitted pair particle has its last collision. The distribution of $r(t_2)^2$
( i.e. of the freeze out points) is displayed in fig. \ref{xxc}. We see that 
this distribution deviates from a
Gaussian form. It can be considered approximately as a sum of two Gaussians.      
\item the {\bf true source radius $R_{true}$ }given by the rms radius of 
the system after the freeze out points have  
been convoluted with a Gaussian wave function squared of a variance of L/4.}
This source radius is called true because it is the relevant source radius if
one wants to determine densities or energy densities and it is also the source
radius the correlation function should "measure". For L=0
$R_{true}= R_{class}$.  
\item the {\bf apparent source radius R} as given by eq.\ref{nneu}. This 
apparent source radius can be calculated from the correlation function 
using the formalism derived in the last section. This is the only radius which
can be measured experimentally.
\end{itemize}

\begin{figure}[hbp]
\centerline{\psfig{figure=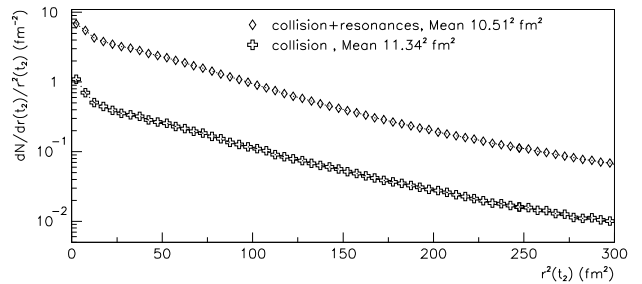,width=8.6cm}}
\caption{Distribution of the freeze out points.}
\label{xxc}
\end{figure}  
We start the investigation by calculating the correlation function 
(eq.\ref{a8} or eq.\ref{a7}) of identically charged pions at midrapidity
for different values of the variance L/4. { We discard for the beginning those
pairs which contain pions coming from resonance decay.} These
correlation functions have a form which can be well described by eq.\ref{nneu}.
We can therefore determine R by fitting the simulation results with a function of
the form of eq.\ref{nneu}.
In fig.\ref{jj1} we display this dependence of the fit parameter $R^2$ on L.
We separate the results 
obtained with (eq.\ref{a7}) and without (eq.\ref{a8}) the smoothness assumption.

Per definition for L=0 both agree. For larger
values they differ by not more than 10\% and hence the smoothness assumption is
acceptable for the analysis of the results of the simulation programs.
For the unrealistic case that $\sqrt{L}$ is large as 
compared to the source we see the expected (see eq.\ref{neun}) linear behavior,
but this is
not true anymore if $\sqrt{L}< R_{class}$, for the 
smoothness assumption as well as for the exact calculation.
\begin{figure}[hbp]
\centerline{\psfig{figure=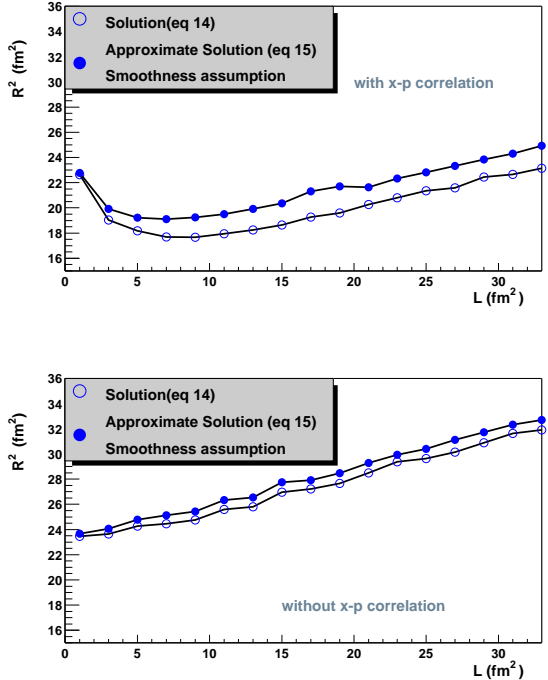,height=11cm}}
\caption{The square of the apparent source radius $R^2$ 
as a function of the variance L/4 of the single particle wave
function squared. 
We display calculations with and without smoothness assumption.
In the bottom
figure we have artificially decorrelated the positions and momenta of the particles.
$R^2$ is obtained by fitting the 
correlation function with  $C(\k)=1+\lambda exp(-4\k^2 R^2) $.}
\label{jj1}
\end{figure}  

In order to understand the origin of this deviation from the expected behavior
we decorrelate position and momentum of the pions. We attribute to each
pion with a given source momentum a  position randomly chosen among the source
positions of the other pions. The result is displayed in the bottom panel of 
fig.\ref{jj1}.  We see now the expected behavior: The apparent source size
increases if the width of the wavefunction and hence the true source radius
increases.

In fig. \ref{jj11} we plot the true source radius as a function of the 
apparent source radius. As explained, both can be
obtained independently in the simulation program. Without space-momentum 
space correlations the apparent source radius follows the true source size,
the convolution of the classical source with the single particle wave function
squared, which can be independently calculated in the simulation program.
The absolute values are, however, different. The reason for this difference is 
that the distribution of emission points (fig. \ref{xxc}) is not a Gaussian 
(despite of the fact
that the correlation function can be well described by a Gaussian). }

If we include the correlations generated in the simulation program
the apparent source radius is, however,  a quite complicated function the true
source radius. Assuming that the apparent source size is the true source size 
one overpredicts the density of the source by up to { a factor of 2}. Thus we conclude 
that for the reaction investigated here the true source
radius cannot even approximately be inferred from the apparent radius without a
detailed knowledge of the space-momentum space correlations . 
We just like to mention that the experiments yield apparent source sizes 
which are considered as too small in order to be the true source size, a fact
which, according to our results, may have its explanation in  
space-momentum space correlations.

\begin{figure}[htbp]
\centerline{\psfig{figure=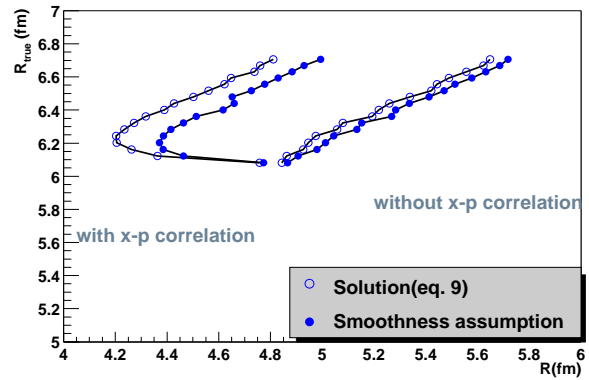,width=8.6cm}}
\caption{The true source size $R_{true}$ (as given by the simulation program) 
as a function of the apparent 
source size R determined by fitting the calculated correlation function by
$C(\k)=1+\lambda exp(-4\k^2 R^2) $ 
including (see fig. \ref{jj1} top) and excluding (see fig. \ref{jj1} bottom) 
space-momentum space correlations.}
\label{jj11}
\end{figure}
We would like to add two remarks:\\  
a) The space-momentum space correlation in the simulation program is not created 
by flow. It is already caused by the string dynamics and is modified 
later due to rescattering. Thus adding only a radial (or 
in plane) flow to models which have otherwise no correlations between space and
momentum space would not reduce the difference between R and $R_{true}$.\\
b) Being forced to introduce L as a free parameter we have lost the 
predictive power of the simulation programs as far as two
body correlations are concerned. Instead of predicting the two body correlation
function we can only state whether there is a value of L for which the 
simulation program reproduces the measured apparent source radius R.
(As said the different values of   
$R_{true}$ are obtained by folding the same classical source distribution with
the square of wave functions with different variances L/4.)

\section{Simulations including electromagnetic interactions}
\subsection{The momentum change due to the electromagnetic interaction}
\begin{figure}[htbp]
\centerline{\psfig{figure=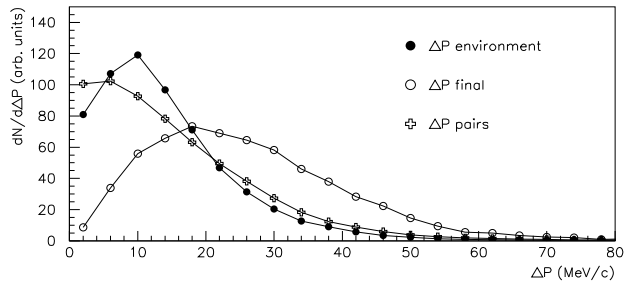,width=\hsize}}
\caption{Distribution of the change of the relative momentum between the
particles of a correlated pair after freeze out. We separate the change 
due to the electromagnetic interaction with the environment
(black circle), with the other pion of the correlated pair (open cross), 
and the sum of both
(open circle). }
\label{momen}
\end{figure}

After freeze out and during their way to the detectors the mesons change
their momentum due to the Coulomb interaction. In fig. \ref{momen}
we display the distribution of the change of the relative momentum of the
correlated pairs between $t_2$ and detection, separated into the different
origins. The average momentum change is of the order of 25 MeV and the
interaction between the correlated pair contributes as much as the interaction
of the pair particles with the environment. Even if these numbers are based on 
classical
trajectory calculations, presented in section II, it is evident that calculations
which take into account only one of these effects are not suited for the
situation at hand.

\subsection{Standard treatment of electromagnetic interactions}

As said already, the correlation function $C(\k)$ differs from one only 
for pairs with a relative momentum $k=\sqrt{(\ka - \kb)^2}  < 100 MeV$.
Only these pairs carry information on the size of the system.   
The change of the relative momentum of a correlated pair due to the Coulomb
interaction is of the same order of magnitude and therefore not negligible.

This is well known since long and several attempts have been made
to correct the two body correlation function for the electromagnetic 
interaction \cite{al,a,b,ac}. 
The correlation function $C(\k, \K)$ for a pair of particles 
emitted with a center of mass momentum $\K$ and 
an asymptotic  relative momentum 
$\k $ is given by the square of the projection of the particle wave function
$\phi^{K}({\bf r})$   
, where $\r$ is the relative distance of the pair 
particles, onto the Coulomb wave function $\Phi^{coul}_{k/2}(\r)$,
\begin{equation}
C(\k, \K)= \int d^3r |<\Phi^{coul}_{k/2}|\r><\r|
\phi^{K}>|^2.
\end{equation}
$\Phi^{coul}_{k/2}(\r)$ is the solution of 
\begin{equation}
( {-\nabla^2_{{\bf r}}\over 2\mu} + V(\r))\Phi^{coul}_{k/2}(\r)= 
 {\k^2\over 2\mu} \Phi^{coul}_{k/2}(\r) .
\label{sch}
\end{equation}

Assuming identical freeze out points for the two particles
$|\phi^{K}(\r)|^2 = \delta^3(\r)$  
we obtain the so called Gamow correction factor
\begin{equation}
G(\eta ) = |\Phi_{k/2}^{coul}(r=0)|^2= \frac{2 \pi \eta}{exp [2 \pi \eta] -1} 
\end{equation}
with $\eta = \mu \alpha / k $ , 
$\mu$ the reduced mass. Often the "Coulomb corrected" correlation function 
at freeze out is calculated by dividing the measured correlation function
by the Gamow correction factor. 
  
 This approach has been recently critically studied and 
extended to a system of n bosons by Alt et al. \cite{ac}. 
They have shown that in a certain limit the n-body wave function can be 
approximated by the product of the relative wave functions of all pairs 
and that for small source sizes ($\approx 1 fm$) this product 
can be replaced by that of the relative wave functions at zero relative 
distance. Then the Coulomb correction is nothing else than a product of Gamow
correction factors. However, already for sources of 5 fm this product of
correction factors overestimates the true Coulomb correction \cite{ac}. 
In view of the large source size we see in neXus (fig.3) these results 
elucidate why the Gamow correction does not work well for heavy ion reactions at
ultrarelativistic energies, as
has been observed in \cite{cs,LED,BRI,FER}.

\subsection{The correlation function in  presence of electromagnetic fields}

The above mentioned quantal approaches for the Coulomb correction 
are implicitly based on the assumption that the correlated pair particles are 
emitted very close to each other. Only then the 
environment changes the momentum of the pair particles by the same amount, 
leaving the relative momentum k unchanged. In this case one can solve 
eq.\ref{sch}. 
This is not the situation we see in the simulations. There
the change of the relative momentum of a correlated pair
particles due to the mutual Coulomb interaction is of the
same order of magnitude as that due to the interaction with all the other 
particles. Then eq. \ref{sch} is not anymore a good approximation even more  
it is not even calculable:
the asymptotic relative momenta k, necessary to calculate the Coulomb wave 
functions, are only known at the very end of the simulation, after the
mutual Coulomb interactions have ceased, and are based in our approach 
on classical trajectory calculations. Thus we are left with the fact, that if
quantal effects are important, our asymptotic relative momenta, based on a
classical calculation are not correct, on the contrary, if they are correct
quantal effects are negligible and we can live with calculating Coulomb
corrections classically, as done here. Thus the quantal Coulomb correction proposed in
\cite{ac} are incompatible with semiclassical simulation programs.

It has been proposed as well to replace in the Coulomb wave function the
asymptotic momentum by the momentum at freeze out. For many applications this is
certainly a very good approximation, but in our context it is not useful: this
ansatz would eliminate the effect we would like to study: How the correlation
function is modified by the interaction of the pair particles with the
environment.    

We would like to mention another problem which renders the application of a static 
quantal Coulomb correction in the context of semiclassical simulation 
programs quite difficult:
Coulomb interactions and strong interactions cannot be separated in time.
The earlier emitted pair particle has moved already in a (strong) Coulomb 
field when the later emitted one has its last collision. 

Due to these problems even the most modest quantal approach, to describe the
correlated pair by a two body wave function in which all other charges are
treated in the Born Oppenheimer approximation, i.e. by supplementing in the
Hamiltonian of eq. \ref{sch}  $V(\r)$ by the Coulomb interaction
with all other charged particles  $\sum_{k }  ( V^C(\r_{ik},t) + V^C(\r_{jk},t)$,
and then to solve the time dependent version of this equation, is not feasible
with present day computers.

What can we do in this situation? Because the exact calculation is not
possible we employ approximative methods to study the influence of
the electromagnetic interaction on the correlation function. These
approximations are crude, too crude probably to yield quantitatively reliable
results, but sufficiently precise to demonstrate that more precise methods have 
to be developed before a quantitative relation between the apparent radius
and the true source radius of the system can be established.

Our approximation
consist in assuming that the momentum transfer due to the electromagnetic
interaction is instantaneous. Under this assumption we study two cases. First we assume that the change 
of the momentum due to the electromagnetic interaction happens immediately at
freeze out. This result is then compared with calculations which assume
that the instantaneous momentum change happens at t, the time where in
neXus on the average half of the momentum change has happened. t is of the 
order of 20 fm/c. Because the momentum change due to the
electromagnetic interaction is moderate it does not completely destroy
the correlation function obtained by neglecting the electromagnetic 
interaction (as for example
collisions with large momentum transfer would do). Rather one expects a smooth
modification. 

The first case is easy to study. We have only to replace \ks \ by \ks +\q \ 
where
\q $= \Delta \k_{pair} + \Delta \k_{environment}$ is the momentum change due to the electromagnetic interaction.
The second case requires a more careful study. As said, in order to create a
correlation function  one has to describe the mesons after freeze out
by a wave function with a finite width in coordinate and momentum space. 
Therefore, if we take the quantum 
character of the mesons seriously, we have now 4 different amplitudes 
which are visualized in fig.9.
\begin{eqnarray*}  
B_1&=& \int d\pa d\pb d\pc d\pd d\pe d\pf \\ 
    &&<\ka,t_2|t_1,\pe>\delta(\pc+\qa-\pe)
    <\pc,t_1|0,\pa>\\
    &&<\pa,0|\Phi_A><\kb,t_2|t_1,\pf>\delta(\pd+\qb-\pf)\\
    &&<\pd,t_1|0,\pb><\pb,0|\Phi_B>\\
    &=&<\ka-\qa,t_1|\Phi_A> <\kb-\qb,t_1|\Phi_B>
\end{eqnarray*}
 
\begin{eqnarray*}  
B_2&=& \int d\pa d\pb d\pc d\pd d\pe d\pf \\ 
    &&<\ka,t_2|t_1,\pf>\delta(\pd+\qb-\pf)
    <\pd,t_1|0,\pa>\\
    &&<\pa,0|\Phi_A><\kb,t_2|t_1,\pe>\delta(\pc+\qa-\pe)\\
    &&<\pc,t_1|0,\pb><\pb,0|\Phi_B>\\
    &=&<\ka-\qb,t_1|\Phi_A> <\kb-\qa,t_1|\Phi_B>
\end{eqnarray*}

\begin{eqnarray*}  
B_3&=& \int d\pa d\pb d\pc d\pd d\pe d\pf \\ 
    &&<\ka,t_2|t_1,\pf>\delta(\pd+\qb-\pf)
    <\pd,t_1|0,\pb>\\
    &&<\pb,0|\Phi_B><\kb,t_2|t_1,\pe>\delta(\pc+\qa-\pe)\\
    &&<\pc,t_1|0,\pa><\pa,0|\Phi_A>\\
    &=&<\ka-\qb,t_1|\Phi_B> <\kb-\qa,t_1|\Phi_A>
\end{eqnarray*}

\begin{eqnarray*}  
B_4&=& \int d\pa d\pb d\pc d\pd d\pe d\pf \\ 
    &&<\ka,t_2|t_1,\pe>\delta(\pc+\qa-\pe)
    <\pc,t_1|0,\pb>\\
    &&<\pb,0|\Phi_B><\kb,t_2|t_1,\pf>\delta(\pd+\qb-\pf)\\
    &&<\pd,t_1|0,\pa><\pa,0|\Phi_A>\\
    &=&<\ka-\qa,t_1|\Phi_B> <\kb-\qb,t_1|\Phi_A>.
\end{eqnarray*}
\begin{figure}[tbp]
\centerline{\psfig{figure=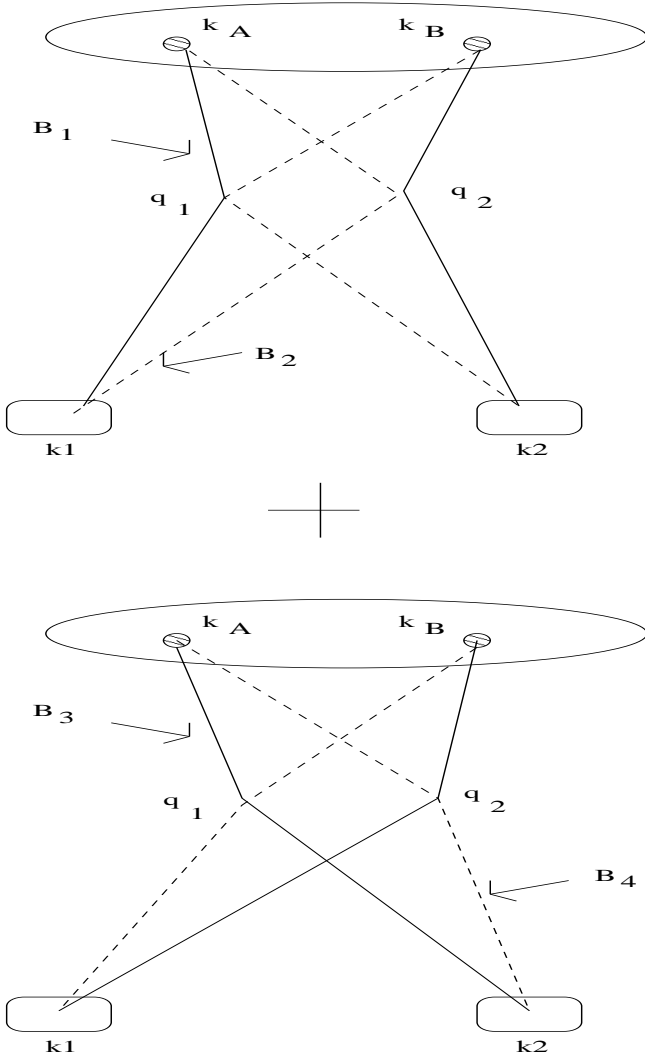,width=8.6cm,height=14cm}}

\caption{The 4 indistinguishable different amplitudes if the electromagnetic interaction happens
instantaneously, with a momentum transfer of $\qb $ and $\qa $ .}
\label{interfer}
\end{figure}
Calculating $|\sum_l B_l|^2$ we find after integration over the center of mass
motion:
\begin{eqnarray*}
&Re(B_1B_2^*) = D\cos(2\q(\rs+2\k\d))e^{-(\k-\ks)^2L -\q^2L}\\
&Re(B_1B_3^*) = D\cos(2\k(\rs+2\q\d))e^{-(\q+\ks)^2L -\k^2L}\\
&Re(B_1B_4^*) = D\cos(2\rs(\k-\q))e^{-(\k-\q)^2L -\ks^2L}\\
&Re(B_2B_3^*) = D\cos(2\rs(\k+\q)))e^{-(\k+\q)^2L -\ks^2L}\\
&Re(B_2B_4^*) = D\cos(2\k(\rs-2\q\d))e^{-(\q-\ks)^2L -\k^2L}\\
&Re(B_3B_4^*) = D\cos(2\q(\rs-2\k\d))e^{-(\k+\ks)^2L -\q^2L}\\
&B_1B_1^* = D e^{-(\k-\q-\ks)^2L}\\
&B_2B_2^* = D e^{-(\k+\q-\ks)^2L}\\
&B_3B_3^* = D e^{-(\k-\q+\ks)^2L}\\
&B_4B_4^* = D e^{-(\k+\q+\ks)^2L}\\
\end{eqnarray*}
where D contains the normalization. The exact result $C(\k) = {|\sum B_l|^2 \over
\sum |B_l|^2}$ is too complex to discuss
the physics. Therefore we apply the smoothness assumption, i.e we assume that 
all exponential functions can be approximated by $e^{-(\k-\q-\ks)^2L}$. Then we
obtain 
\begin{eqnarray}
C_{SA}(\k) &=&
(\sum_{s=1}^{N_{ev}}\sum_{i=1}^{N_{pairs(s)}}e^{-(\k-\q-\ks_{s,i})^2L} \nonumber
\\
&&(1 + 0.5 \times e^{-(\k-\q)^2 L} \; cos(2\rs_{s,i}(\k -\q))+ \nonumber \\
&& 0.5 \times e^{-(\k+\q)^2 L} \; cos(2\rs_{s,i}(\k +\q))+ \nonumber \\
&& e^{-\k^2 L} \; cos(2\rs_{s,i}\k)cos(4\q\k\d) + \nonumber\\
           &&  e^{-\q^2 L} \; cos(2\rs_{s,i}\q)cos(4\q\k\d)) \nonumber \\
           &&/(\sum_{s=1}^{N_{ev}}\sum_{i=1}^{N_{pairs(s)}}
            e^{-(\k-\q-\ks_{s,i})^2L}).
\label{tra}
\end{eqnarray}

If \k \ is equal 0 the
correlation function is not anymore two but $\propto 1+e^{-\q^2} \; cos(2 \r\q)
$.  
For finite \k \ and if  \q  \ is small we find 
$C_{SA}(\k)= 2 (1 + e^{-\k^2 L}cos(2\k\rs))$ and hence up to a normalization constant the 
same result as in section IV. It is important to realize that 
the change of the momentum due to the 
electromagnetic forces modifies the result in a nonlinear way. 

This general result will be compared later with that obtained under the
assumption that \qa \ is the momentum transfer of the particle emitted with the
momentum \kA \ (and correspondingly that the particle emitted with \kB \ suffers a
momentum transfer of \qb).

\subsection{The correlation function in neXus including electromagnetic 
interactions}

\begin{figure}[htbp]
\centerline{\psfig{figure=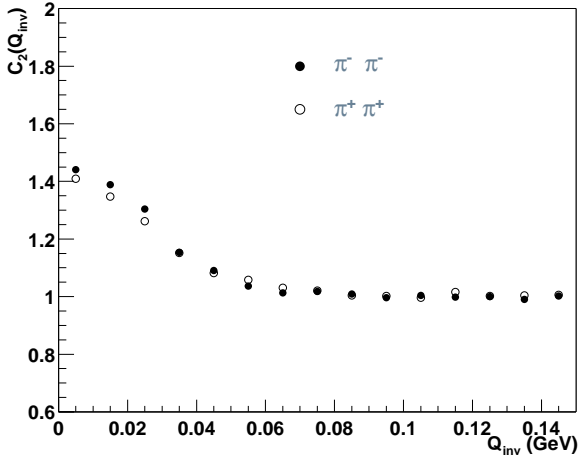,width=8.6cm}}
\caption{Correlation functions for $\pi^+\pi^+$ and $\pi^-\pi^-$ pairs emitted
at midrapidity in 158 GeV/N Pb+Pb central collisions} 
\label{corpipm}
\end{figure}

Before discussing our results in detail we compare the calculated correlation
function for $\pi^+\pi^+$ and $\pi^-\pi^-$ correlations in fig.\ref{corpipm}.
Here we have assumed that the change of the momentum due to the electromagnetic
interaction happens at $t_2$ and we choose $L=0 fm^2$.  
As observed experimentally
(fig. \ref{pipi}) the correlation function is almost identical for  
$\pi^+$ and $\pi^-$ pairs even including the Coulomb interaction in 
distinction to the single particle spectra (fig. \ref{pipi}). Please note
that (as discussed in section IV A) the correlation function would be strongly 
modified if one uses mixed
events for the normalization. In true simulation events the correlated pair 
particles suffer from a mutual Coulomb interaction before freeze out, 
whereas in mixed events this not the case.

We present now our results for three different scenarios. All are
based on the assumption that the momentum transfer due to the Coulomb potential
is instantaneous. In the first scenario we assume
that the change of the momentum due to the electromagnetic
interaction happens at $t_2$, in the second and third scenario that it 
happens at the time when in the classical trajectory calculation 
on the average half of the momentum change due to
the electromagnetic interaction
has occurred. The latter scenario is  calculated for two different
assumptions. First we assume that the particle emitted at A suffers a momentum
change of ${\bf q_1}$ (and that emitted at B changed its momentum by ${\bf q_2}$).
 Then only the trajectories $B_1$ and $B_3$ of fig.\ref{interfer} are allowed. 
Finally we admit all 4 amplitudes $B_1-B_4$.

\begin{figure}[htbp]
\centerline{\psfig{figure=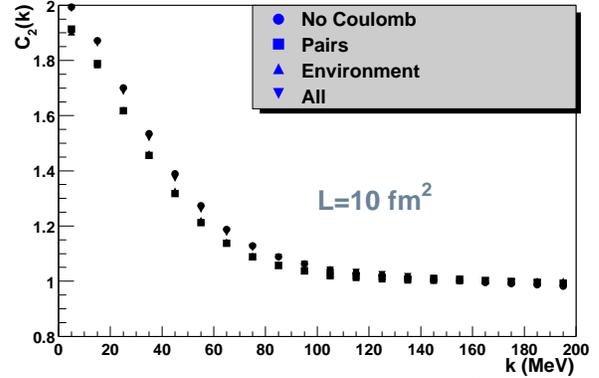,width=8.6cm}}
\caption{ 
Correlation functions for 4 different final momenta (see text). We assume 
that the momentum transfer due to electromagnetic interaction happens at 
freeze out. L is assumed to be = $10fm^2$. } 
\label{mback}
\end{figure}

Fig.\ref{mback} displays the correlation function under the assumption that the
momentum transfer happens immediately at freeze out. We separate  4 
different source momenta to discriminate the different origins of momentum
change: a)$\ks$ momentum at freeze out (No Coulomb) 
b)$\ks$ + ${\bf \Delta k_{environ}}$ (Environment), the change of
the relative momentum of the pair due to the interaction with the environment.
c)$\ks$ + ${\bf \Delta k_{pair}}$ (Pairs), the change of
the relative momentum of the pair particles due to their mutual Coulomb interaction and
d)${\bf k_{final}}$=$\ks$ + ${\bf \Delta k_{pair}}$ + ${\bf \Delta k_{environ}}$ (All).
The correlation functions are rather similar and the fit yields very similar
results for R. Due to the fact that we don't mix events for the determination 
of the denominator of eq.\ref{a8}, 
this result was expected. Event mixing would create a shift of the maximum to a
finite value of $Q_{int}$ due to the lack of Coulomb repulsion between the particles
from different events.

\begin{figure}[htbp]
\centerline{\psfig{figure=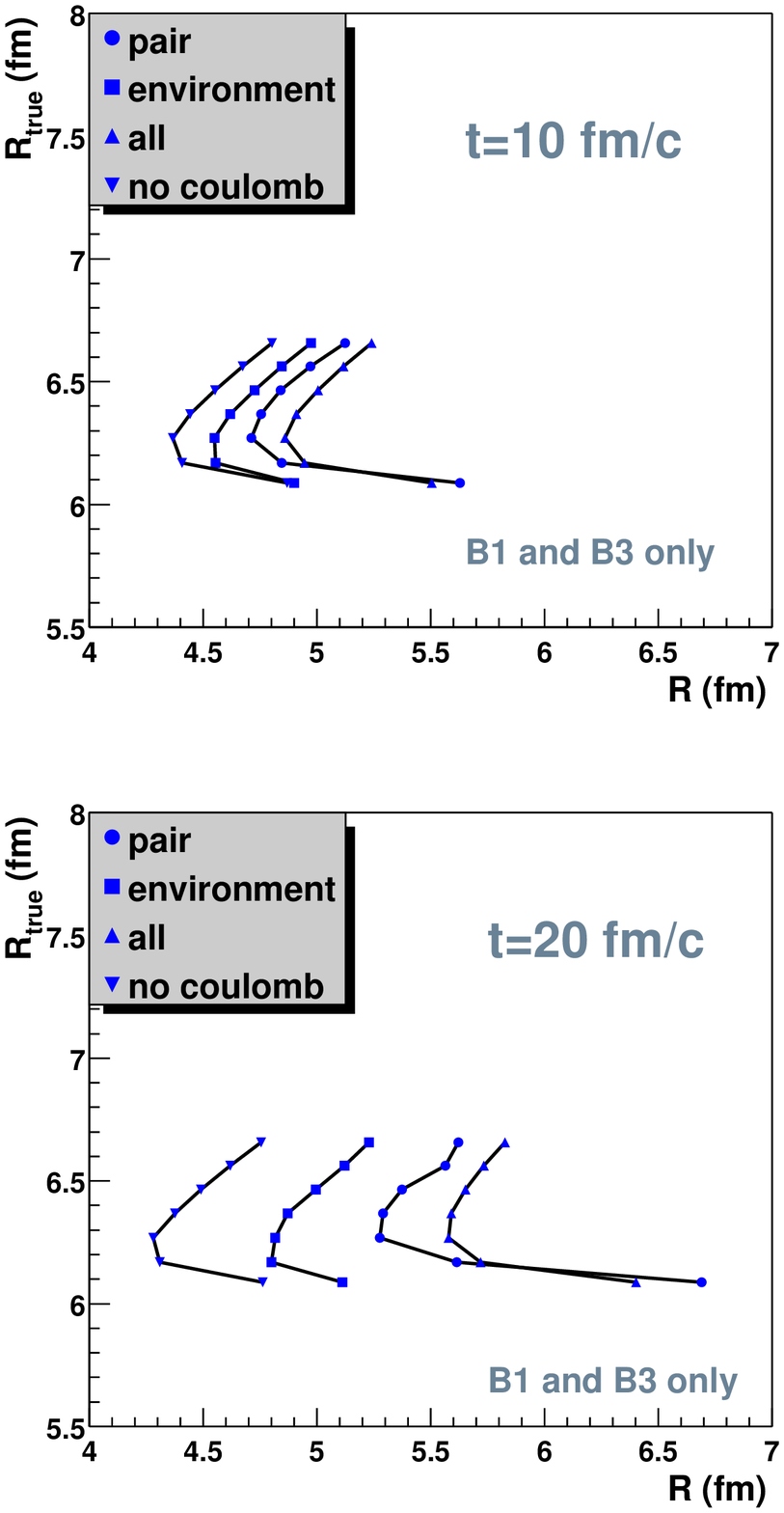,width=8.6cm}}
\caption{
The true source radius $R_{true}$ (as given by the simulation program) 
as a function of the apparent source radius R determined by
$C({\bf k })=1+\lambda exp(-{4\bf k}^2 R^2) $ including the Coulomb interaction
(see text).
We admit here only the amplitudes 
$B_1$ and $B_3$ (see text) and assume that the momentum transfer due to the
electromagnetic
interaction takes place at t= 10 fm/c (left) or at 20 fm/c (right). }
\label{c10}
\end{figure}

If we assume that the momentum transfer happens at 10 fm/c resp.
20 fm/c after $t_2$ and admitting only the
amplitudes $B_1$ and $B_3$ we observe the dependence of the true source 
on the apparent source radius R as displayed in fig.\ref{c10}. { For a given
source size $R_{true}$ the Coulomb interaction increases
the measurable apparent source size. The later the
momentum transfer happens the larger is this change. For a given $R_{true}$ 
a momentum transfer at 20 fm/c may increase of the apparent source size radius 
by as much as 25\%. }No Coulomb is identical with the curve
displayed in fig.\ref{jj11}. We like to mention that the change of the apparent
source radius due to the Coulomb interaction is of the order of magnitude
expected from the schematic study in ref. \cite{gb}.

\begin{figure}[htbp]
\centerline{\psfig{figure=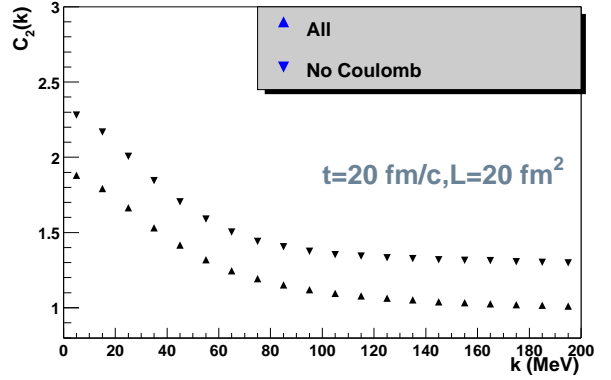,width=8.6cm}}
\caption{
Correlation function obtained for the general case where all amplitudes 
$B_1 - B_4$ (see text) are admitted. We assume 
that the momentum transfer takes place at t= 20 fm/c. The correlation function
is not anymore Gaussian. 
 }
\label{c++bd}
\end{figure}

If we allow for all 4 different amplitudes $B_1 -B_4$ we see another time a quite
different behavior. The correlation function has not anymore a Gaussian form
as can be seen in fig.\ref{c++bd}. Fitting only the low momentum component  by a
Gaussian function one obtains R values which are
slightly lower than if only the amplitudes $B_1$ and $B_3$ are allowed. The error bars of the fit are, however, large. We would
like to mention that the non Gaussian form of the correlation function becomes
only evident in our approach. If the correlation function is normalized by
event mixing an overall normalization appears. Because the correlation function
is normalized to 1 at large values of k this overall normalization absorbs
the large \k \  offset seen in fig.\ref{c++bd}.    

\begin{figure}[htbp]
\centerline{\psfig{figure=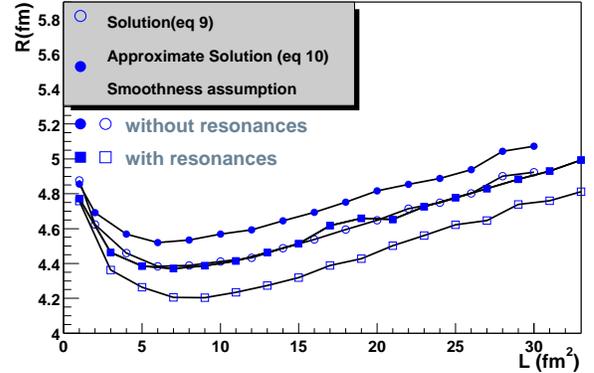,width=8.6cm}}
\caption{The apparent source radius R determined by
$C({\bf k })=1+\lambda exp(-{4\bf k}^2 R^2) $ as a function of L including and excluding the 
pairs in which at least one pion comes from resonance decay }
\label{c++re}
\end{figure}

The influence of the pions from resonance decay on the correlation function 
is rather weak { although the resonances can decay very late, when 
the distance to its pair partner is very large. This large distance
increases $R_{true}$ tremendously. Therefore we have excluded them up to 
now from the analysis.  The apparent source radius of all pairs as compared to pairs of
particle where no pion comes from resonance decay is displayed  
in fig.\ref{c++re} as a function of L (L/4 being the variance of the wave
function).}  
This result is understandable: at the late time when the resonances disintegrate
there is no other particle close by and hence the value of $\q\r(t_2)$
is large. On the average this contribution is therefore negligible.
\section{correlation function of the kaons}
\begin{figure}[htbp]
\centerline{\psfig{figure=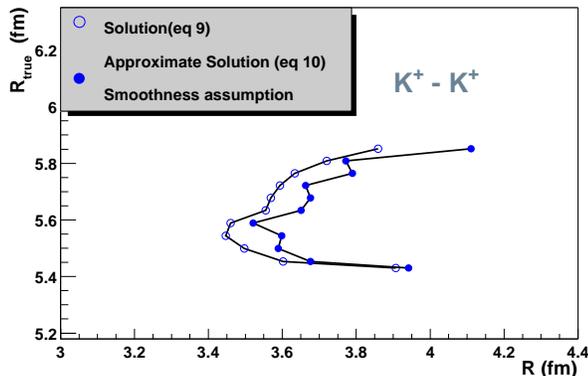,width=8.6cm}}
\caption{The true source radius $R_{true}$ (as given by the simulation program) 
for kaons as a function of the apparent source radius R determined by
$C({\bf k })=1+\lambda exp(-{4\bf k}^2 R^2) $.}
\label{c++k}
\end{figure}

So far we have investigated pions. In the simulation programs we
have sufficient $K^+$ to allow for the calculation of a 
correlation function. The result is displayed in fig.\ref{c++k}.
We observe a very similar behavior as for the pions because the 
space-momentum space correlations are similar. The kaon source is smaller than
that of the pions: $R_{true}$ as well as  R are
smaller. The $K^+$ cross sections are lower and 
therefore the freeze out happens earlier. 
Consequently the smaller source radius is
a natural consequence of known physics. 
We observe as well a large discrepancy between $R_{true}$  and R.

\section{Conclusion}

We studied in detail two problems which one encounters if one tries to
interpret HBT correlation functions obtained from numerical simulations 
of heavy ion reactions at CERN energies: The influence of space - momentum space correlations and 
the influence of final state Coulomb interactions. For this purpose we 
supplemented one of the presently available programs, neXus, 
by electromagnetic interactions.  

We find, first of all, that there is nothing like a static source which emits
particles. The particles come from sources of quite different
sizes because the system expands as a whole. 
Momenta and positions of the particles are strongly correlated and hence the 
source is not chaotic. Both observations render the task rather difficult to 
extract from the correlation function useful information on the size of the 
system or on densities. The most one can expect to obtain are a time averaged 
values. However, even these time averaged quantities are neither unique 
(different particle species give different values) nor representative (pairs 
of particles with small relative momenta (to which the correlation function is 
sensitive) are emitted from a smaller source then arbitrary pairs).

In order to calculate a correlation function in these semiclassical simulation
programs one has to introduce (at least) one free parameter. Here we assume that
the particles have a Gaussian wave function after their last collisions. 
Consequently, the simulation programs cannot predict the two body correlation 
functions. They can only determine for which variance the calculated 
two body correlation functions agrees with experiment.

Knowing the variance of the wave function and the freeze out points we can
calculate in the simulation programs directly the average rms radius of the 
source. It is called $R_{true}$. All correlation functions we obtain have the 
form $C(\k) = 1 + \lambda e^{-4\k^2R^2}$, the same form as seen in the 
experiments. Thus the crucial problem
is to relate the apparent source radius R with physically meaningful quantities
like $R_{true}$.

The simulation program tells us that there is no simple relation between
R and $R_{true}$ due to four different reasons:
  
1) The distribution
of the emission points is not Gaussian, whereas (due to the the finite error
bars in the simulation as well as in the experiments) the correlation function
is well described by a Gaussian and finer details of the source cannot be 
resolved in the correlation function. Therefore there 
is a difference between R and $R_{true}$ already of the order of 20\%. 
Furthermore,
the distribution of emission points and the distribution of sources (as
displayed in fig.3) differ as well considerably. The latter is the relevant
quantity if one would like to extract densities or energy densities.

2) For the reaction investigated the strong space-momentum space correlations 
modify the relation between R and  $R_{true}$ in an important way. Thus R does 
not measured the true size of the source but underestimates it by $\approx 16
\%$ and hence the true average density of the system is $60\%$ lower than that
extract from R. This may explain the
fact that the experimental source sizes, when analyzed under the assumption that
the source is (almost) chaotic are incomprehensibly small. The relation 
between R and  $R_{true}$ depends in a complicated way on the dynamics of the
system and is not due to quantities like flow. Thus it is not evident how the
information on space-momentum space correlations can be incorporated in more
phenomenological approaches and, consequently, how this type of models can
extract the true source radius from the observed apparent source radius.    

3) The Coulomb interaction changes  the relation between R and 
$R_{true}$ as well. Assuming that our calculation gives quantitatively reliable results, the
difference between R and  $R_{true}$ can reach $30\%$ and hence the densities
differ by a factor of 2.
For the influence of the Coulomb interaction we presented here only very
approximative results. They are based on classical Coulomb trajectory
calculations. They show that the Coulomb interaction between the 
correlated pairs and the other charges is as important as the Coulomb
interaction between the particles of the correlated pair and  
that the standard procedure 
to use Gamow correction factors is not justified, because the average particle
distance at freeze out is too large.  

4) The difference between R and  $R_{true}$  depends on the particle species
because their cross section are different. 

Thus this study shows that two problems have to be solved before the
correlation function measurements can provide useful information on the density
of the system. We have a) to understand how the apparent source radius is related
to the time averaged true source radius and b) how the true source radius can
be related to physical quantities like energy densities or densities. Here we
have addressed the first question only and found that the difference between 
R and  $R_{true}$ is all but negligible. Hence a naive interpretation of the
observed apparent source radius as the radius of the system at freeze out is
certainly premature.

The Coulomb interaction produces still another effect if one employs the
standard method of event mixing because the absence of the Coulomb 
interaction in mixed events modifies the correlation function. This effect we
have observed in the simulations but not discussed here. It will further add
to the difficulty to relate the apparent source radius to the true source size.

\acknowledgments
We would like to thanks K. Werner , H.J Drescher for helping us with the neXus
code . Interesting discussions with P. Braun-Munzinger, L. Conin , U. Heinz, 
S. Pratt, F. Reti\`ere, 
K. Werner and U. Wiedemann are gratefully acknowledged.


\begin{references}

\bibitem{GGLP} G. Goldhaber, S. Goldhaber , W. Lee and A. Pais , Phys. Rev {\bf
}, 300 (1960)
\bibitem{HBT56} R. Hanbury-Brown and R.Q. Twiss, Nature {\bf178 }(1956) 1046 
\bibitem{ccs} T. Cs\"org\"o et B. L\"orstad, Phys. Rev. {\bf C 54} (1996) 1390
and Nucl. Phys. {\bf A590} (1995) 465c  . 
\bibitem{HEINZ} U. A. Wiedemann , U. Heinz Phys.Rept.{\bf 319} (1999) 145
\bibitem{gy} K.Kolehmainen and M. Guylassy, Phys. Lett. {\bf B180} (1986) 203,
S.S. Padula and M. Guylassy, Nucl. Phys. {\bf A498} (1989) 555c and
Phys. Lett {\bf B217} (1989) 181 
\bibitem{cso}T. Cs\"org\"o et al., Phys. Rev. Lett. {\bf B 241} (1990) 301
\bibitem{pr1} S. Pratt, T. Cs\"org\"o and J. Zimanyi Phys. Rev. {\bf C42} (1990)
2646 
\bibitem{sul} J.P. Sullivan at al., Phys. Rev. Lett. {\bf70} (1993) 3000\\
Nucl. Phys. {\bf A566} (1994) 531c, \\
T.J.Humanic et al, Phys. Rev. {\bf  C53} (1996) 901 
\bibitem{gb}G. Baym and P. Braun-Munzinger Nucl. Phys {\bf A610} (1996) 286c 
\bibitem{al} E.O. Alt et al., Eur. Phys. J. {\bf C13}:663, (2000)
\bibitem{a} R. Lednicky, V.L. Lyuboshitz, Sov. J. Nucl. Phys. {\bf 35}: 770
(1982)
\bibitem{b} R. Lednicky,V.L. Lyuboshitz,B. Erazmus and D. Nouais, Phys. Lett.
{\bf B373}, 30 (1996)
\bibitem{ac}E. O. Alt, T. Cs\"org\"o, B. Lorstad, J. Schmidt-Sorensen, proceedings
of  30th International Symposium on Multiparticle Dynamics (ISMD 2000), 
Tihany, Hungary (World Scientific, Singapore in press) and
E. O. Alt, T. Cs\"org\"o, B. Lorstad, J. Schmidt-Sorensen, Eur.Phys.J.{\bf C13} 
(2000)663
\bibitem{AIC} J. Aichelin Nucl. Phys A {\bf 617} (1997) 510
\bibitem{WER93}  K. Werner Phys.Rept. {\bf 232}:87,(1993)
\bibitem{W1} H. J. Drescher, M. Hladik, S. Ostapchenko,T.Pierog  K. Werner 
preprint  hep-ph/0007198 (2000)
\bibitem{URQ98} S.A. Bass, M. Belkacem, M. Bleicher, M. Brandstetter, L.
 Bravina, C. Ernst, 
L. Gerland, M. Hofmann, S. Hofmann, J. Konopka, G. Mao, L. Neise, S. Soff, C.
Spieles, H. Weber,
L.A. Winckelmann, H. St\"ocker, W. Greiner, C. Hartnack, J. Aichelin
, N. Amelin Prog.Part.Nucl.Phys.{\bf 41 }:225,(1998)
\bibitem{pratt} Y.D. Kim et al, PRC {\bf 45} , 387(1992)
\bibitem{BOGG}  F. Reti\`ere, PhD thesis, University of Nantes, 2000 , see also for  NA44 :H. Boggild et al Phys. Lett B {\bf 372}:339,(1996),
\bibitem{zhang} Q.H. Zhang et al., Phys. Lett {\bf B 407}:33 (1997)
\bibitem{sc} Scott Pratt, Phys. Rev. {\bf C56}, 1095 (1997)
\bibitem{cs}S. Pratt, T. Cs\"org\"o and J. Zimanyi, Phys. Rev. {\bf C42}:2646 (1990)
\bibitem{LED} Yu.M. Sinyukov, R. Lednicky, X. V. Akkelin, J. Pluta and B. Erazmus
Phys.Lett. {\bf B432}:248-257 (1998)
\bibitem{BRI} D. Brinkmann , Thesis, IKF, University of Frankfurt.
\bibitem{FER} Na35-Collaboration: D. Ferenc and al Z. Phys {\bf C7  }: 443-448
(1997)
\end{references}
\end{document}